\documentclass[%
 aip,
 amsmath,amssymb,
 reprint,%
 floatfix,
]{revtex4-1}
\usepackage{graphicx}
\usepackage{dcolumn}
\usepackage{bm}
\usepackage{float}
\usepackage[utf8]{inputenc}
\usepackage[T1]{fontenc}
\usepackage{mathptmx}
\usepackage{etoolbox}
\usepackage{ulem}
\usepackage{xcolor}
\usepackage{soul,xcolor}
\usepackage{upgreek}
\makeatletter
\def\@email#1#2{%
 \endgroup
 \patchcmd{\titleblock@produce}
  {\frontmatter@RRAPformat}
  {\frontmatter@RRAPformat{\produce@RRAP{*#1\href{mailto:#2}{#2}}}\frontmatter@RRAPformat}
  {}{}
}%
\makeatother
\begin{document}

\preprint{RSI23-AR-02313}
\title[]{Tuning Methods for Multigap Drift Tube Linacs}
\author{O. Shelbaya}
 \email{oshelb@triumf.ca.}
 \affiliation{TRIUMF,
             4004 Wesbrook Mall,
             Vancouver,
             V6T 2A3,
             BC,
             Canada}
\affiliation{Department of Physics and Astronomy, University of Victoria,
             PO Box 1700 STN CSC,
             Victoria,
            V8W 2Y2,
             BC,
             Canada}
\author{R. Baartman}%
 \affiliation{TRIUMF,
             4004 Wesbrook Mall,
             Vancouver,
             V6T 2A3,
             BC,
             Canada}
 \affiliation{Department of Physics and Astronomy, University of Victoria,
             PO Box 1700 STN CSC,
             Victoria,
            V8W 2Y2,
             BC,
             Canada}
\author{P.\,Braun}
 \affiliation{Institute for Applied Physics, Goethe Universit\"at,
             Max-von-Laue-Straße 1,
             D-60438,
             Frankfurt a.~M.,
             Germany}
\author{P.\,M.\,Jung}
 \affiliation{TRIUMF,
             4004 Wesbrook Mall,
             Vancouver,
             V6T 2A3,
             BC,
             Canada}
 \affiliation{Department of Physics and Astronomy, University of Victoria,
             PO Box 1700 STN CSC,
             Victoria,
            V8W 2Y2,
             BC,
             Canada}   
\author{O. Kester}
 \affiliation{TRIUMF,
             4004 Wesbrook Mall,
             Vancouver,
             V6T 2A3,
             BC,
             Canada}
 \affiliation{Department of Physics and Astronomy, University of Victoria,
             PO Box 1700 STN CSC,
             Victoria,
            V8W 2Y2,
             BC,
             Canada}   
\author{T. Planche}%
 \affiliation{TRIUMF,
             4004 Wesbrook Mall,
             Vancouver,
             V6T 2A3,
             BC,
             Canada}
 \affiliation{Department of Physics and Astronomy, University of Victoria,
             PO Box 1700 STN CSC,
             Victoria,
            V8W 2Y2,
             BC,
             Canada}
\author{H.\,Podlech}
 \affiliation{Institute for Applied Physics, Goethe Universit\"at,
             Max-von-Laue-Straße 1,
             D-60438,
             Frankfurt a.~M.,
             Germany}
\affiliation{Helmholtz Research Academy Hesse for FAIR
(HFHF), 60438 Frankfurt a. M, Germany}
\author{S.\,D.\,R\"adel}
 \affiliation{TRIUMF,
             4004 Wesbrook Mall,
             Vancouver,
             V6T 2A3,
             BC,
             Canada}

\date{\today}

\setstcolor{red}
\begin{abstract}
Multigap cavities are used extensively in linear accelerators to achieve velocities up to a few percent of the speed of light, driving nuclear physics research around the world. 
Unlike single-gap structures, there is no closed-form expression to calculate the output beam parameters from the cavity voltage and phase. To overcome this, we propose to use a method based on the integration of the first and second moments of the beam distribution through the axially symmetric time-dependent fields of the cavity. 
A beam-based calibration between the model's electric field scaling and the machine's rf amplitudes is presented, yielding a fast on-line energy change method, returning cavity amplitude and phase necessary for a desired output beam energy and energy spread. The method is validated with $^{23}$Na$^{6+}$ beam energy measurements. 
\end{abstract}

\maketitle

\section{Variable Energy Linear Accelerators}



\begin{figure}[!b]
    \centering
    \includegraphics[width=0.47\textwidth]{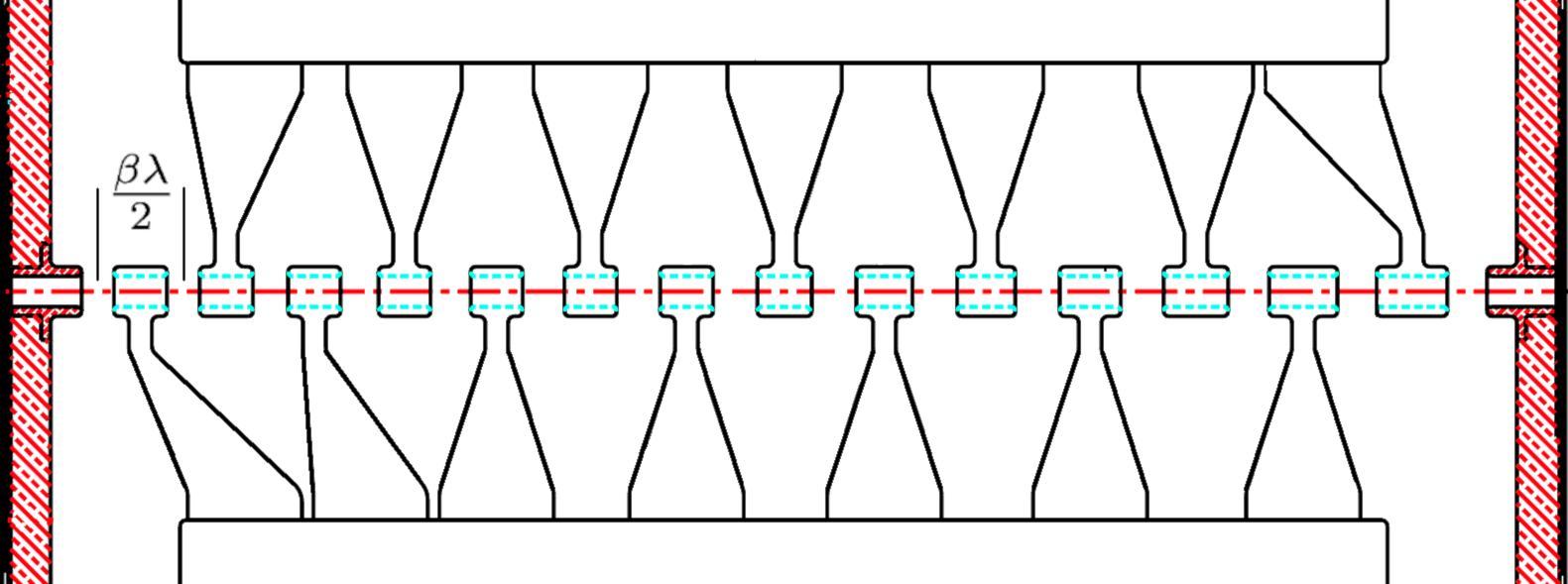}
    \caption{Cross-sectional drawing of the third IH accelerating cavity in TRIUMF ISAC-DTL, operating at with $f$\,=\,106\,MHz, consisting of 15 gaps about an axis of radial symmetry (red, dotted line). Each drift tube is held in place by a stem (digit). For scale, $\beta\lambda/2=4.5$\,cm}
    \label{fig:tank1}
\end{figure}

Particle accelerator research facilities around the world, such as REX-ISOLDE\cite{kester2003accelerated}, GSI\cite{Barth:2000bf} and TRIUMF\cite{ball2020triumf} have been pursuing investigations into the properties of nuclei across the nuclear chart, using multigap rf cavities to achieve requisite beam energies. Using interdigital H-Type structure (IH) cavities\cite{blewett1956linear,Gerigk:2011ph}, such as shown in the cross-section drawing in Fig.\,\ref{fig:tank1}, these can be operated at variable output\cite{laxdal1997separated} beam velocity making them an ubiquitous choice in ion linacs. The relatively small drift tubes enable a high shunt impedance for a variety of rf phase and amplitude combinations\cite{ratzinger2019combined}.

Separated function\cite{laxdal1997separated} machines enable shorter rf cavities when compared the Alvarez type linac, enabling larger energy gain in a shorter physical length. Flexibility in the IH structure's cell length permits design of low energy reference particle acceleration where the relative change in $\beta$\,$=$\,$v/c$ can be appreciable within even a single tank\cite{Xiao:2019xlq}. With such structures, as we are about to see, the relationship between the output energy and the cavity amplitude and phase can no longer be described by a simple cosine function with a constant `transit time factor' (TTF)\cite{panofsky1951linear}.

\begin{figure}[!htpb]
    \centering
    \includegraphics[width=0.49\textwidth]{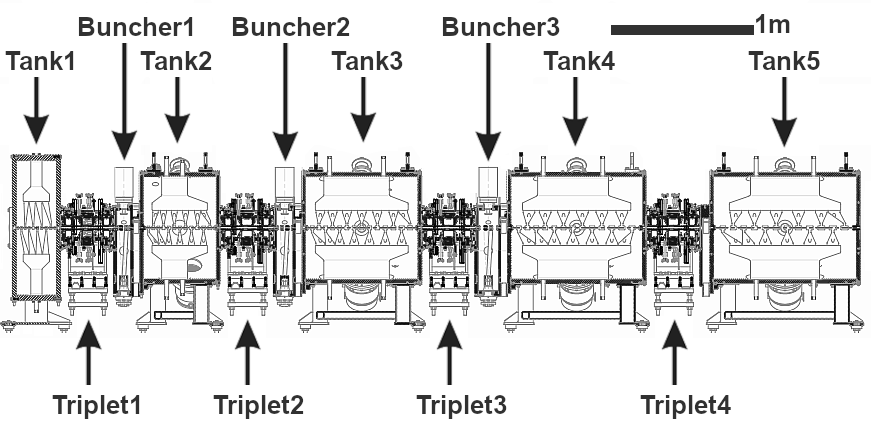}
    \caption{Schematic representation of the ISAC-DTL, consisting of five IH accelerating tanks and three longitudinal bunching cavities. Quadrupole triplets provide transverse focusing.}
    \label{fig:DTL}
\end{figure}

To calculate the cavity amplitude and phase required to produce a desired output velocity and energy spread, we propose a method that combines the accuracy of a multiparticle simulation with the speed of an envelope code\cite{Shelbaya:2021sth}: Since a single reference particle is tracked through the fields, together with the second moments of the distribution centered around it, this results in a more computationally lightweight and therefore faster runtime, when compared to ray-tracing\cite{shelbaya2019fast}. After exposing the method, we present its application to the case of TRIUMF's ISAC-DTL\cite{Dutto:2001wy} (Figure \ref{fig:DTL}), a separated function drift tube linac\cite{laxdal1997separated} based upon the combined zero degree structure (KONUS)\cite{ratzinger2019combined,Laxdal:2000hx}. A beam-based calibration, obtained by performing scans of the output energies vs.\ rf phase, is presented. With this, it is possible to use constrained optimization to compute DTL rf settings on-line, and rapidly change the output energy of the structure. The simulation's predicted reference particle energy is compared to on-line measurements of an $^{23}$Na$^{6+}$ beam to validate the approach. This can be applied to any DTL, given knowledge of its on-axis electric field, obtained either by bead-pulling measurements or simulations.

\section{Method}

\subsection{Panofsky Equation-Based Methodology}
The energy gained by a charged particle passing through a single rf gap can in general be written as~\cite{panofsky1951linear,Wangler:2008zz}:
\begin{equation}
    \Delta E = qV_0T\cos\theta,
    \label{eq:energyTTF}
\end{equation}
where $q$ is the particle charge, $\theta$ is its phase with respect to the gap midpoint, $V_0$ is the integral of the on-axis longitudinal electric field known as the effective voltage or amplitude voltage, and $T\in[0,1]$ is the `transit-time factor' for a single gap of length $L$, with rf wavelength $\lambda$ and reference particle velocity $\beta c$\cite{Wangler:2008zz}:
\begin{equation}
    T = \frac{\sin(\pi L / (\beta\lambda))}{\pi L/(\beta\lambda)}.
    \label{eq:singleGapTTF}
\end{equation}
The value of $T$ depends on the particle velocity and gap geometry, and is independent of $\theta$ in the particular case where all of the following conditions are met: (i) the origin of time $t$\,=\,0 is chosen at the center of the gap, (ii) the cavity on-axis field profile $\mathcal{E}(s)$ is an even function, and (iii) the velocity change of the particle across the gap is negligible\cite{Wangler:2008zz}. 

The elegant simplicity of this approach is lost when a particle traverses a multigap structure and sees its velocity change significantly. Though multigap structures can be modelled as a sequence of individual gaps, each with its own $T$, its own $\theta$ and its own $V_0$, this computation is more involved: The particle velocity and time of arrival at each gap is not known {\it a priori}, and must be calculated numerically. Consequently, the value of both the transit time factor and phase parameter for every gap is a function of the cavity amplitude and phase setting. The resulting cavity output energy versus rf amplitude and phase can be measured and compared to machine simulations\cite{Lallement:2017uxp,shelbaya2021autofocus} or represented as a Fourier series whose amplitude coefficients are fit using on-line measurements\cite{Marchetto:2008zz}.
Since numerical integration is required in any case, it is advisable to abandon~eq.~\ref{eq:energyTTF}, and instead evaluate the longitudinal equations of motion through the actual on-axis field of the cavity.

\subsection{Proposed Method}
As outlined in previous work\cite{Baartman:2017faa,shelbaya2021autofocus}, we choose to use the Frenet-Serret coordinate $s$, which tracks the arclength of the reference particle's trajectory, as independent variable. Then, the state vector of any particle in the beam can be written as:
\begin{equation}
    {\bf X} = (x,P_x,y,P_y,z,\delta P)
\end{equation}
where the longitudinal canonical coordinates, ordinarily time and energy deviations, have been scaled by the synchronous particle velocity $\beta_0c$ so that they are homogenous to a length and a momentum, while preserving their canonical nature. 
\begin{equation}
    z  = \delta t\beta_0 c = 
    \text{, \ and \ } z' \equiv \delta P = \frac{\delta E}{\beta_0 c}.
\end{equation}

In {\sc transoptr}, the bunch length $z$ is fundamentally a time-spread scaled by the reference particle velocity so that it is recorded in units of length for consistency with the transverse bunch dimensions ($x,y$). All three ($x,y,z$) are differentials with respect to the reference particle. The longitudinal coordinates are differentials with respect to the reference particles coordinates $t_0$ and $E_0$. These are found by integrating the two equations derived from the Hamiltonian $H_s$
\begin{equation}
\frac{{\rm d}E_0}{{\rm d}s} = \frac{\partial H_s}{\partial t}\mbox{, \ and \ }
\frac{{\rm d}t_0}{{\rm d}s} = -\frac{\partial H_s}{\partial E} = \frac{E_0}{P_0} = \frac{1}{\beta_0 c}.
\label{eq:energytime}
\end{equation}
In the idealized case with centred beam and correctly chosen Frenet-Serret reference trajectory, forces are linear and particle motion equations are
\begin{align}\label{eq:eom1}
    \frac{\text{d}\mathbf{X}}{\text{d}s} & =\mathbf{F}\mathbf{X}
\end{align} 
where the 6x6 matrix $\mathbf{F}$ is derived from the quadratic terms in the Hamiltonian as $\mathbf{F}=\mathbf{J}\mathbf{H}({H_s})$, $\mathbf{J}$ is the elementary symplectic matrix and $\mathbf{H}({H_s})$ is the Hessian matrix of the Courant-Snyder Hamiltonian\cite{Courant:1958wbj}. 

The beam centroid is represented by the first moments:
\begin{equation}
    \langle\mathbf{X}\rangle=\frac{1}{N}\sum_{i=1}^N \mathbf{X}\,,
\end{equation}
where $N$ is the total number of particles, and is zero in the idealized case when there are no forces on the beam centroid away from the reference trajectory. If there are none, the system Hamiltonian, expanded to second order, contains no first order terms. In the real case, there is broken symmetry resulting from unknown misalignments of focusing elements. Moreover, there can be transverse rf forces arising e.g.\ from the drift tube support stems in the DTL. Representing the extra transverse forces by the vector $\mathbf{F}_\perp$, the equation of motion for the beam centroid is:
\begin{equation}\label{eq:eomc}
    \frac{\text{d} \langle\mathbf{X}\rangle}{\text{d} s} =\mathbf{F}\langle\mathbf{X}\rangle+\mathbf{F}_\perp .
\end{equation} 
Note that this represents only four equations as the longitudinal centroid is zero by eqn.\,\ref{eq:energytime}. Further, note that this is only valid for cases where the bunch center and synchronous particles coincide longitudinally.

The beam's rms size and statistical correlations are the elements of the covariance matrix, also known as the beam matrix\cite{brown1967slac}:
\begin{equation}
    \mathbf{\sigma}=\frac{1}{N}\sum_{i=1}^N (\mathbf{X}-\langle\mathbf{X}\rangle)(\mathbf{X}-\langle\mathbf{X}\rangle)^\text{T}.
\end{equation}
For simple cases where the transfer matrix $\mathbf{M}$ is known analytically, this matrix transforms as\cite{brown1967slac} \begin{equation}\label{eq:matr}\boldsymbol{\sigma}_{\rm f}=\mathbf{M\boldsymbol{\sigma}_{\rm i} M^{\rm T}}.\end{equation} For the general cases, though, the equations of motion for the second moments can be written as\cite{sacherer1971rms}:
\begin{equation}
    \label{eq:eom2}
    \frac{\text{d} \mathbf{\sigma}}{\text{d} s} =\mathbf{F}\mathbf{\sigma} + \mathbf{\sigma}\mathbf{F}^\text{T} .
\end{equation} There are 21 equations as the covariance matrix is symmetric. Adding eqns.\,\ref{eq:energytime},\ref{eq:eomc} there are thus 27 first order equations. These are solved simultaneously in the code {\sc transoptr}\cite{heighway1981transoptr,baartman2016transoptr} using a Runge-Kutta technique. This naturally allows for the evaluation of the continuous transit efficiency through the field:
\begin{equation}
    T(s) = \frac{\int\mathcal{E}(s)\cos(\omega t(s) + \phi){\rm d}s}{\int\mathcal{E}(s){\rm d}s} .
    \label{eq:tofs}
\end{equation}
The parameter $\phi$ is a constant: It is the absolute rf phase of the field and is independent from the particle dynamics or gap geometry. It is analogous to the rf parameter that is adjusted during cavity operation. The spatial on-axis field profile $\mathcal{E}(s)$ of the cavity has been obtained either by bead-pulling measurement or finite-element simulation. Likewise, the continuous field weighted synchronous phase can be defined:
\begin{equation}
\phi_w(s) = \frac{\int\mathcal{E}(s)(\omega t(s) + \phi){\rm d}s}{\int\mathcal{E}(s){\rm d}s},
\label{eq:phase}
\end{equation} which is the analogue of the gap crossing phase, for a continuous integration through the field. We apply this to low intensity rare ion beams\cite{shelbaya2019fast,Shelbaya:2022eyc}, but in the general case, with space charge, the $\mathbf{F}$ matrix is augmented with the usual space charge elements\cite{sacherer1971rms} in eq.~\ref{eq:eom2} (but not in eq.~\ref{eq:eomc}, as a consequence of Newton's third law; image charges are ignored).

A limitation of this technique is that the nonlinearity of the sinusoidal rf cannot be taken into account. Though the reference particle longitudinal coordinates are correctly tracked, the motion of the other particles with respect to it only use the linearized rf forces. Bunches with $\omega\delta t>\sim \pi$/4 or in the cases considered below, $z>\sim 1$\,cm, cannot be accurately represented. For most cases, though, where longitudinal rms emittance is intended to be conserved, the optimal operation regime is within this constraint\cite{shelbaya2019fast}.

\section{Variable Output Energy}
\begin{figure}[!t]
    \centering
    \includegraphics[width=0.47\textwidth]{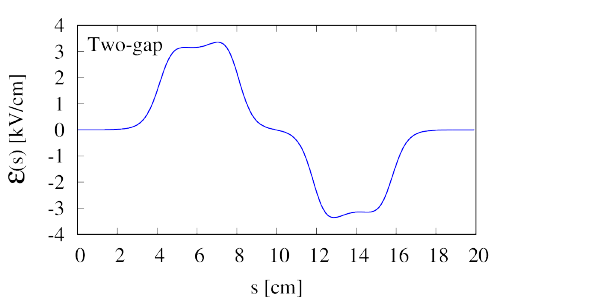}\\
    \includegraphics[width=0.47\textwidth]{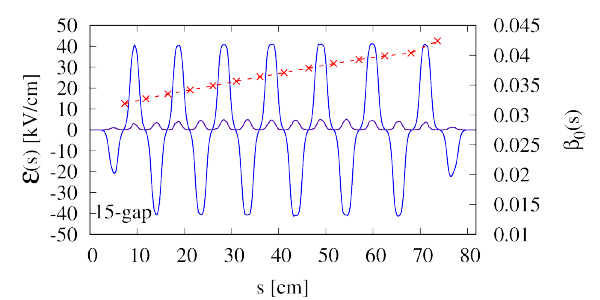}\\ 
    \includegraphics[width=0.47\textwidth]{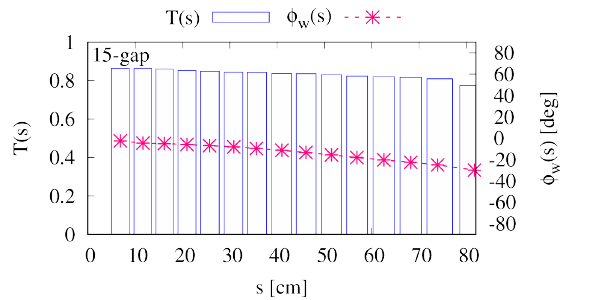}
    \caption{\label{fig:twofields}Top: on-axis electric field intensity for a 35\,MHz two-gap room temperature resonator, located upstream of the ISAC-DTL\cite{mitra199935}, operated at injected $\beta\lambda$\,$=$\,17\,cm. This cavity is used to time-focus beams into the first DTL IH tank. Middle: on-axis electric field intensity (blue) for an 106\,MHz 15 gap IH structure (Fig.\,\ref{fig:DTL}, Tank-3), where beams are injected at $\beta\lambda$\,$=$\,10\,cm. For the IH structure, the design synchronous particle velocity is shown in red. A dipole component (purple) is due to the support stems (see Fig.\,\ref{fig:tank1}). Bottom: the transit time factor and synchronous phase for the 15-gap resonator, at its design tune. Fields generated\cite{shelbaya2020transoptr,TRI-BN-19-02} in the code {\sc opera2d}\cite{fields1999opera} and {\sc cst-mws}\cite{studio2008cst}, respectively.}
\end{figure}

Multiple gaps used for acceleration have their cell length $l_c$ designed to produce the desired $\beta_0(s)$ profile. For an rf wavelength $\lambda$, the synchronous velocity through the field relates to the Widerøe condition as:
\begin{equation}
    l_c(s) = \frac{\beta_0(s)\lambda}{2}
    \label{eq:cellLength}
\end{equation} (See Fig.\,\ref{fig:tank1}.)
To reach a different output energy than the design value, structures that operate at a fixed frequency must be operated in a de-tuned rf configuration. The phase and scaling must produce the required beam velocity while also minimizing longitudinal divergence. 
Two cases are considered in this section, both of which are shown in Fig.\,\ref{fig:twofields}: A symmetric two-gap structure and a 15-gap IH structure where the reference velocity changes by more than 30\%. To illustrate this method, we choose the tank which possesses the most gaps in the ISAC-DTL: Tank-3. In the first example the Panofsky equation~\ref{eq:energyTTF} is a very good approximation; in the second case 
the energy as function of cavity phase and amplitude is far more complex and does not follow a simple cosine relation. At the bottom of the figure, the design transit efficiency from eq.~\ref{eq:tofs} and synchronous phase of eq.~\ref{eq:phase} are shown for the 15-gap cavity, showing both phase slippage and the variation of $T(s)$ along the structure. The envelope code {\sc transoptr} is used for the analysis in both cases. 
\subsection{Two-Gap Example\label{sec:twogap}}
  A 2-gap device is sufficiently short that, provided the fractional energy gain is small, it can be understood as a thin lens. The energy gain eq.\,\ref{eq:energyTTF} varies along a bunch and so can be differentiated to find the imparted energy spread:
  \begin{equation}
  \frac{d}{d\phi}\Delta E=-qV\sin\phi\mbox{, or, }\frac{d}{d\phi}\frac{\delta p}{p}=-\frac{qV}{2E}\sin\phi.
  \end{equation}
  As $d\phi=-\frac{2\pi}{\beta\lambda}d z$, let us integrate in the thin lens approximation:
  \begin{equation}\label{wof}
  \Delta\frac{\delta p}{p}=\Delta z'=\left(\pi\frac{qV}{E}\frac{\sin\phi}{\beta\lambda}\right)\Delta z=-\frac{\Delta z}{f}.
  \end{equation}
The quantity in parentheses can be thought of as a `longitudinal focal power' $-1/f$, as the longitudinal coordinates $(z,z')$ transform according to the transfer matrix $\mathbf{M}=\begin{pmatrix}1&0\\-1/f&1\end{pmatrix}$. Applied to the longitudinal part of the $\boldsymbol{\sigma}$-matrix $\boldsymbol{\sigma}=\begin{pmatrix}\sigma_{55}&\sigma_{56}\\\sigma_{56}&\sigma_{66}\end{pmatrix}$, the transformed matrix is (\ref{eq:matr})
\begin{equation}
\mathbf{M\boldsymbol{\sigma} M^{\rm T}}=\begin{pmatrix}\sigma_{55}&\sigma_{56}-\frac{\sigma_{55}}{f}\\\sigma_{56}-\frac{\sigma_{55}}{f}&\sigma_{66}-2\frac{\sigma_{56}}{f}+\frac{\sigma_{55}}{f^2}\end{pmatrix}.
\end{equation}
As function of $1/f$, this is a minimum when \begin{equation}\label{wof2}
\frac{1}{f}=\frac{\sigma_{56}}{\sigma_{55}}=\frac{\tilde{z'}}{\tilde{z}}\,r_{56},
\end{equation}
where $r_{56}$ is the longitudinal correlation parameter, and the tilde symbols $({\tilde{z}},{\tilde{z'}})$ denote 2rms values of bunch length and momentum spread.

\begin{figure}[!t]
    \centering
    \includegraphics[width=0.48\textwidth]{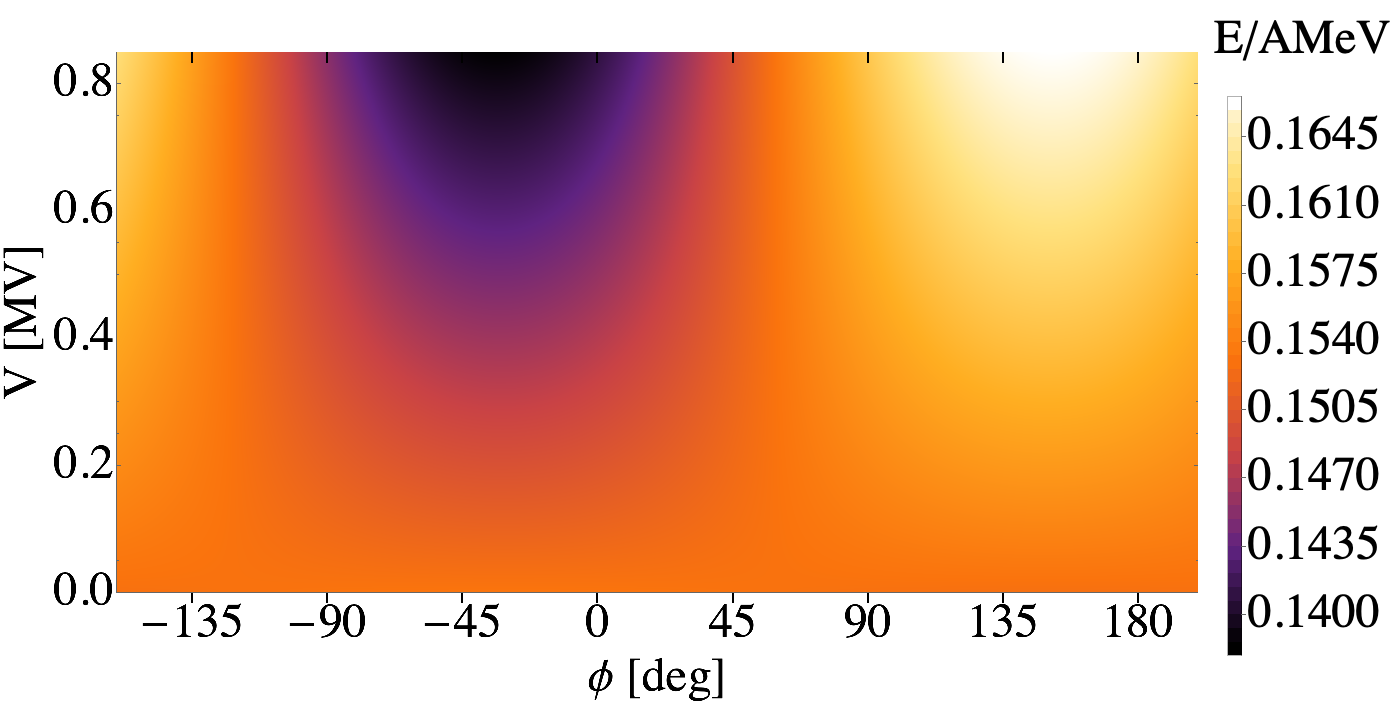}
    \includegraphics[width=0.48\textwidth]{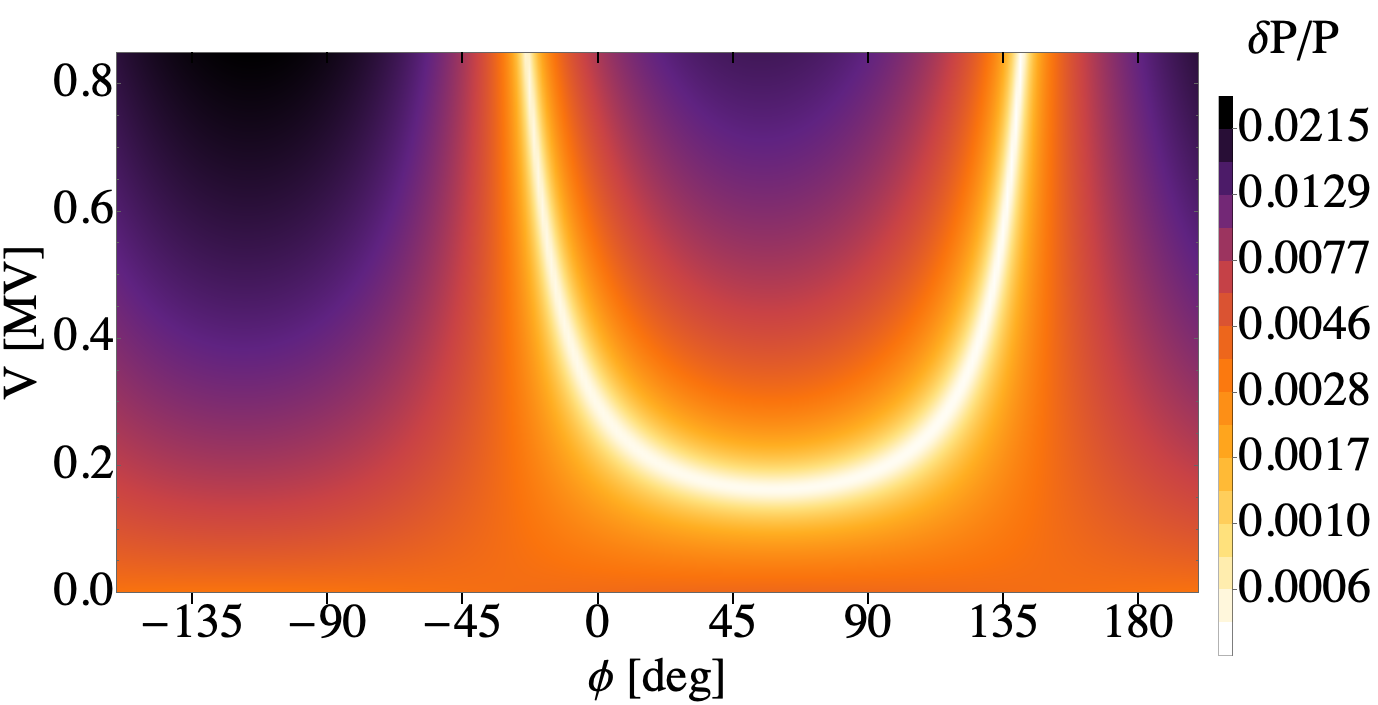}
    \caption{\label{fig:topo1}{\sc transoptr} parameter scans for the two-gap cavity, with $\mathcal{E}(s)$ shown at the top of Fig.\,\ref{fig:twofields}, using an $A$\,=\,30 beam which is longitudinally diverging at cavity injection. An injected longitudinal distribution spanning roughly $\pi/4$ radians, with a momentum spread of 0.2\% is used to illustrate the longitudinal output beam dynamics\cite{shelbaya2021autofocus}. The scans show output $E/A$ in MeV/u and $\delta P/P$. Minimized values of the latter produce optimized energy gain solutions, defined by eq.\,\ref{eq:debunch2gap}.}
\end{figure}

The point to make is that for any combination of desired energy gain and bunch shape outcome, a unique pair for $V$ and phase $\phi$ can be determined analytically. For example, if the desire is that the resulting bunch be in a ``debunched state'' with minimum energy spread, we find combining eqns.\ \ref{wof} and \ref{wof2}:
\begin{equation}\label{eq:debunch2gap}
qV=-2E\frac{\beta\lambda}{2\pi f}\csc\phi=-2E\frac{\beta\lambda}{2\pi\tilde{z}}\tilde{\frac{\delta p}{p}}r_{56}\csc\phi
\end{equation}
This cosecant dependence is evident in the amplitude-phase parameter map shown in Fig.\,\ref{fig:topo1}. Note that the on-crest phase here is $154^\circ$ rather than zero, so the minimum of $-\csc\phi$ is at $64^\circ$ rather than $-90^\circ$. This figure shows {\sc transoptr} envelope simulations of an $A=30$\,u, where u denotes atomic mass units, beam after transit through the two-gap field. At the top of the figure, the reference particle $E/A$ is shown against rf ($\phi,V$) while the bottom shows $\delta P/P$\,=\,$\sqrt{\gamma_z\epsilon_z}$, where the latter are the standard longitudinal Twiss or Courant-Snyder parameters. Since the reference velocity in the field changes by less than 2\%, both output energy and momentum can be understood using the transit-time method, without evaluating any equations of motion.

\subsection{Fifteen-Gap Example}

Unlike the two-gap case, the multigap structure's ($\phi,V$) optima cannot be straightforwardly solved as per eq.\,\ref{eq:debunch2gap}. Approximations exist, as shown in Wangler's textbook\cite{Wangler:2008zz}, however these assume a constant synchronous phase through the field. But, the multiple gap crossings and the considerable resulting change in $\beta_0(s)$ (Fig.\,\ref{fig:twofields}, middle, red) produce an $E/A$ output at optimum phasing which is nonlinear in $V$. Phase slippage in the tank decouples the global rf phase $\phi$ and the individual phases of each rf gap. This causes the emergence of a multitude of oscillatory reference particle $E/A$ solutions through the field, gaining and losing energy during transit. 

\begin{figure}[!t]
    \centering
    \includegraphics[width=0.47\textwidth]{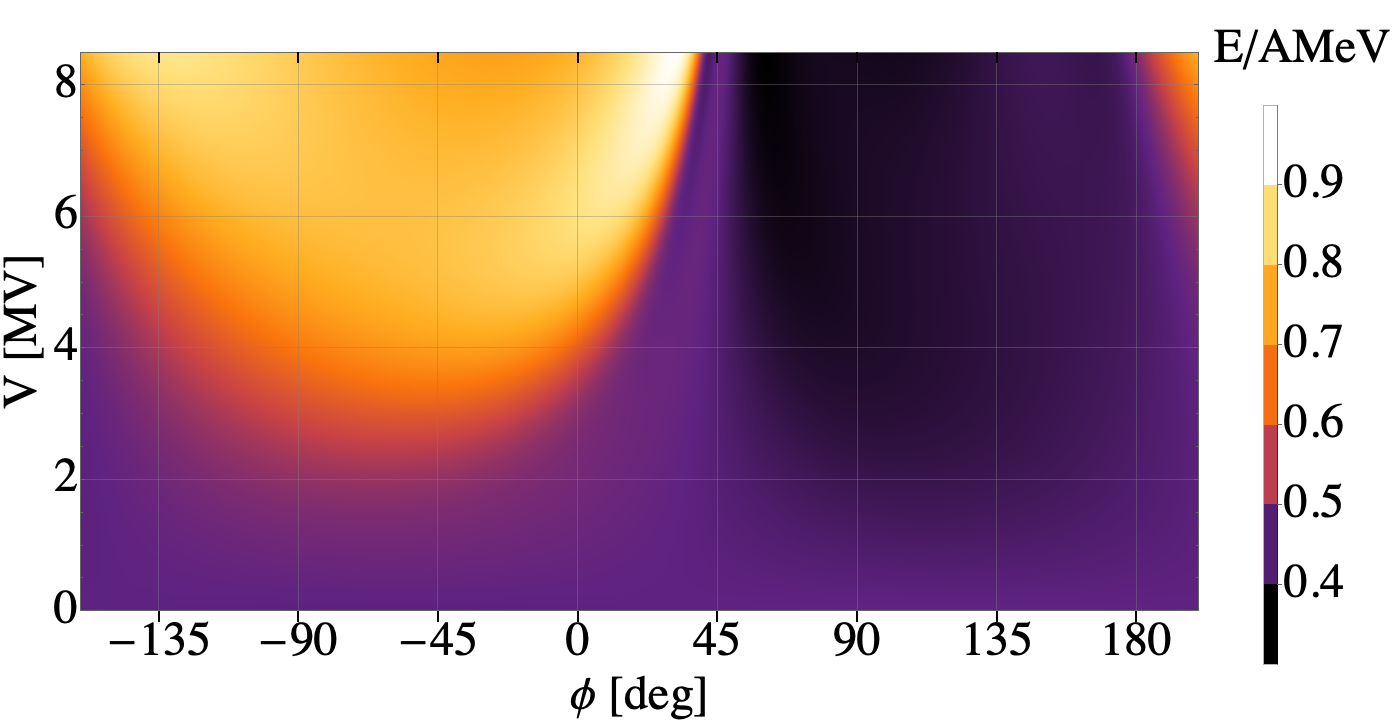}
    \includegraphics[width=0.47\textwidth]{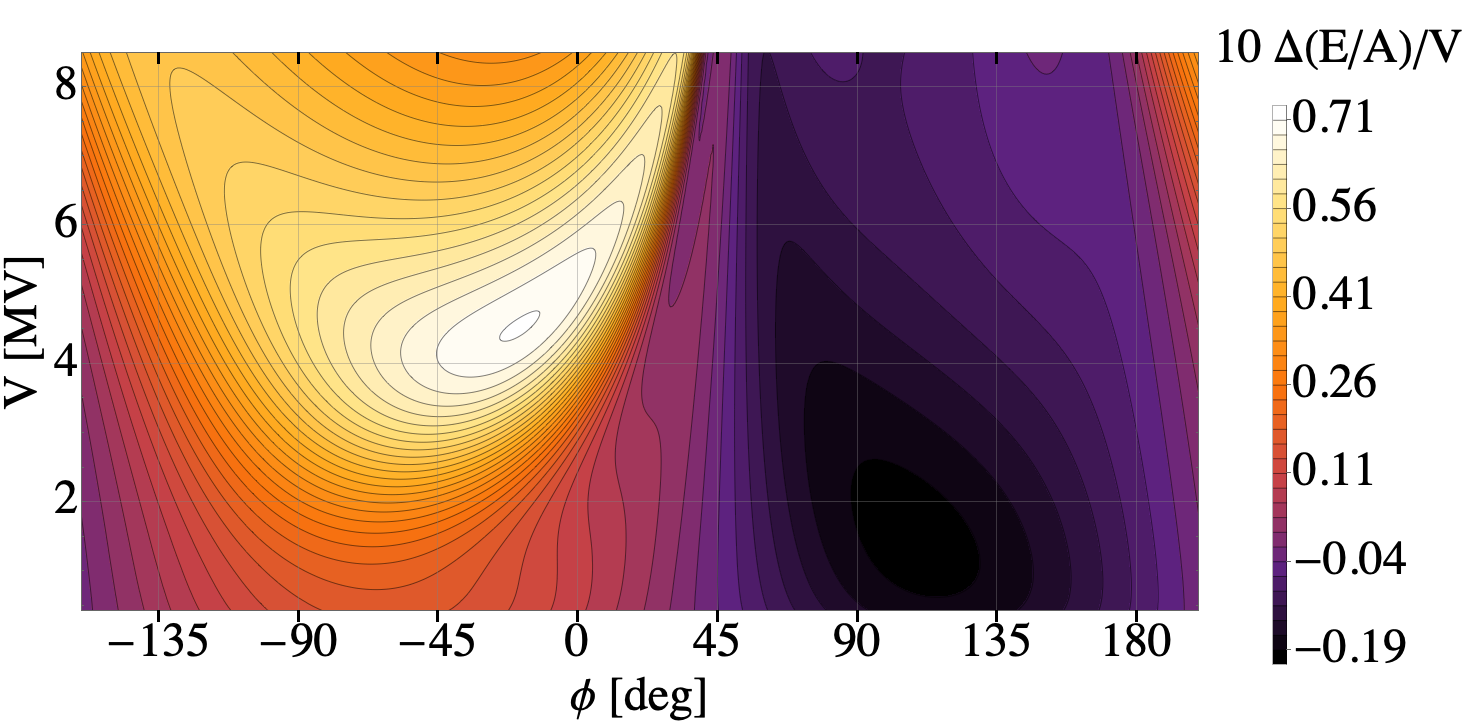}
    \includegraphics[width=0.47\textwidth]{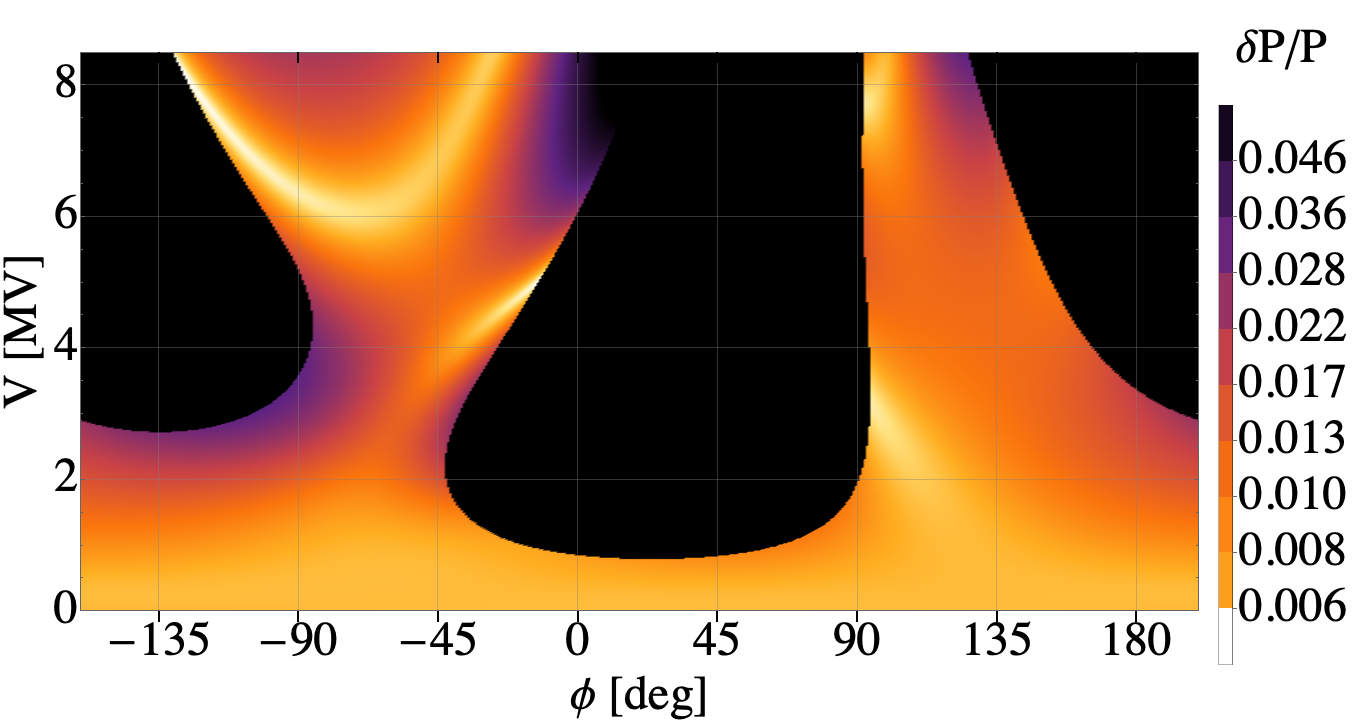}
    \caption{\label{fig:topo}{\sc transoptr} parameter scans for the 15-gap IH structure, with $\mathcal{E}(s)$ shown in the middle of Fig.\,\ref{fig:twofields}, using an $A$\,=\,30 beam. An injected bunch of 2rms length $\tilde{z}=6$\,mm ($24^\circ$), with a momentum spread of $\tilde{z'}=0.5$\% is used to illustrate the longitudinal output beam dynamics\cite{shelbaya2021autofocus}. The top scan shows the output energy per mass unit, and the middle plot shows the same, but divided by the amplitude $V$.  The bottom plot shows $\tilde{z'}=\delta P/P$. The light colour tracks indicate loci of good beam quality. The blackened regions are those where the 2rms bunch length exceeds 1\,cm.} 
\end{figure}

This is shown in Figure~\ref{fig:topo}, with output $E/A$ at the top. It being far from obvious where the design ($\phi,V$) are, the middle plot shows the energy gain divided by $V$. This pinpoints the region $\sim(-26^\circ,\sim5\,\mbox{MV})$ for which the tank was designed, where every gap crossing is close to being ``on crest''. 

In the bottom of Figure~\ref{fig:topo}, the momentum spread is plotted. The {\sc transoptr} calculation solves the exact linear differential equations, but does not take account of the nonlinearity of the rf waveform, and thus large parts of the parameter space are invalid and have been blacked out. These regions are where the bunch is too long and the linear part of the dynamics does not dominate the energy spread. At the design phase and amplitude, one can see that the momentum spread is near a minimum value. Notice however, there is also a crescent-shaped region of good longitudinal beam quality at higher amplitude and more negative phase. 

\begin{figure}[!htpb]
    \centering
     \includegraphics[width=0.5\textwidth]{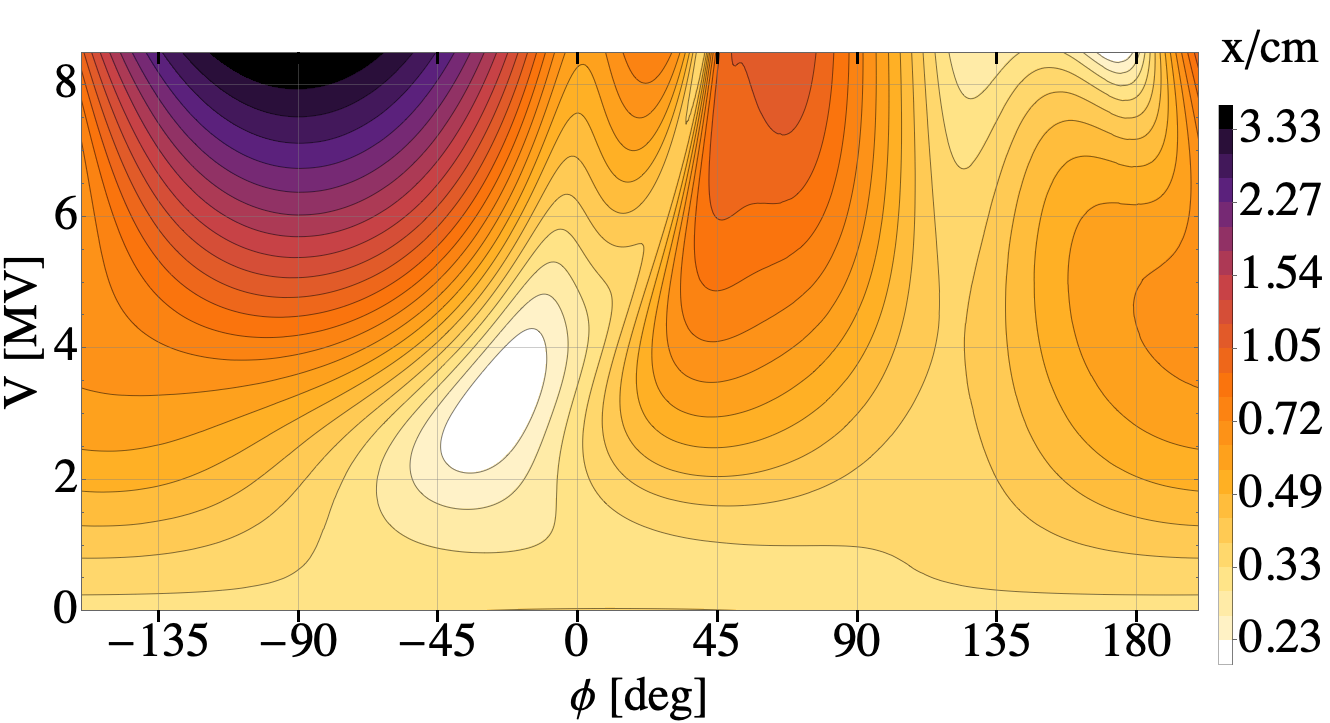}
    \caption{\label{fig:transverse}{\sc transoptr} parameter scans for the 15-gap IH structure, showing transverse beam size at the exit.}
\end{figure}

    \label{eq:phiofs}
The superior longitudinal focusing arises from phases that are bunching in nature (trailing particles gain more energy), but as one would expect, it results in strong transverse defocusing. The existing triplets would not be able to compensate for the strong transverse focus of this case. On the other hand, the 5\,MV case results in an exit beam size equal to the entrance size, but diverging instead of converging. This is exhibited in Figure \ref{fig:transverse}, where the effects of rf defocusing are weakest in the locus of optimum energy gain, since the bunch is subjected to minimal restoring forces\cite{shelbaya2021autofocus}.

\begin{figure}[!t]
    \centering
     \includegraphics[width=0.5\textwidth]{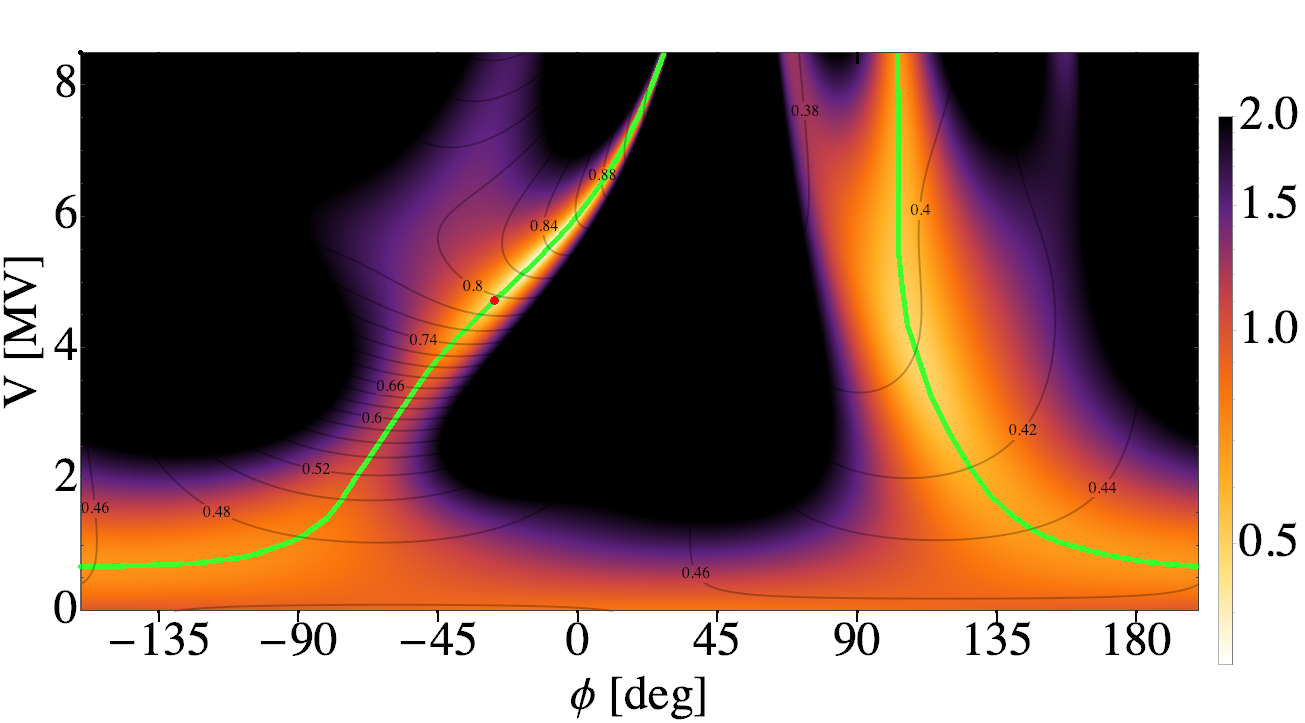}
    \caption{\label{fig:x+z}{\sc transoptr} parameter scans for the 15-gap IH structure, showing graded colour for the quadrature sum of 2rms transverse beam size and 2rms bunch length at the exit. The colour bar is in units of cm. This has been plotted with contours of output energy. The contours are in units of MeV/u. The green track represents phase-amplitude combinations that result in best output beam quality. Beam is injected at 0.46\,MeV/u, and can be decelerated to 0.40,MeV/u or accelerated to 0.95\,MeV/u, though the cavity can be powered only to the design value represented by the red point.}
\end{figure}

The two effects can be combined by plotting the quadrature sum of bunch length and transverse size. This is shown in Fig.\,\ref{fig:x+z}. The green line shows the locus of all acceptable beam quality (long. and transverse), for various output energy. The contours of energy are shown as well. Contours are in increments of 20\,keV/u. The green track starting at the upper right produces output energy 0.40\,MeV/u (decelerated from the injected 0.46\,MeV/u); proceeding down and to the right, it goes off the right edge at the case where it maintains energy and so is bunching without acceleration, continues at the bottom left edge, proceeds up through the design case of $-26^\circ$ and $\sim5$\,MV (0.8MeV/u). Beyond this design amplitude, at the top of the graph at $+30^\circ$ phase it reaches energy $0.95$\,MeV/u (though this is well beyond the capability of the installed power amplifier).

\begin{figure}[!htpb]
    \centering
    \includegraphics[width=0.48\textwidth]{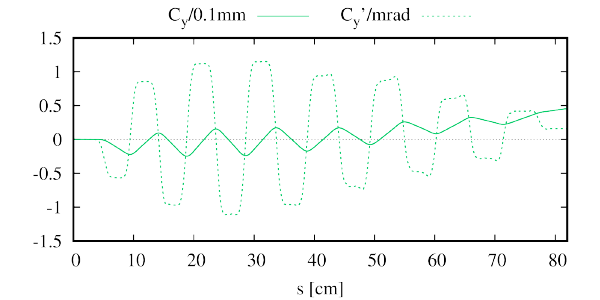}
    \caption{\label{fig:dipoleKicks}{\sc transoptr} computed vertical ($y$) reference particle centroid displacement $C_y$ (solid line) and transverse divergence $C_y'$ (dotted line) in the 15-gap IH structure, at optimum energy gain.}
\end{figure}
\begin{figure}[!htpb]
    \centering
    \includegraphics[width=0.48\textwidth]{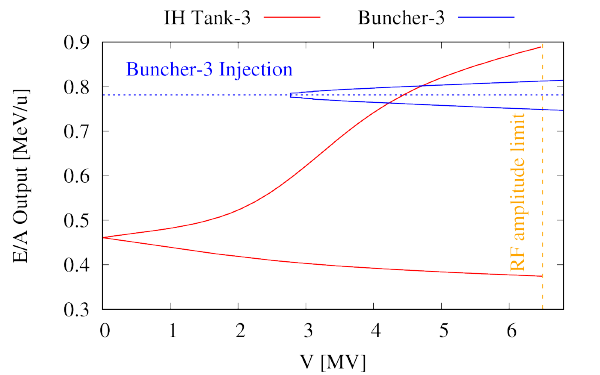}
    \caption{\label{fig:ramp}Output $E/A$ for a given field scaling $V$, where {\sc transoptr} has optimized $\phi$ for minimized $\delta P/P$ at the cavity exit. This is the operating configuration of Tank-3, corresponding to minima in $\delta P/P$ at the bottom of Figure \ref{fig:topo}. The typical rf amplifier limit is shown.}
\end{figure}

One must also consider the expected vertical kicks to the reference particle due to the dipole electric field introduced by the support stems (see Fig.\,\ref{fig:tank1}). Are these sufficient to affect transmission? This effect has been evaluated in {\sc transoptr}, using the vertical electric field in the cavity, shown in the middle of Fig.\,\ref{fig:twofields}. The maximum vertical displacement is found to be roughly 50\,$\upmu$m at the cavity exit, with the vertical centroid and its divergence shown in Figure \ref{fig:dipoleKicks}. For reference, this produces a vertical centroid displacement of roughly 5\,mm after a 10\,m drift and can easily be corrected with vertical magnetic steerers along the beamline. This is consistent with previously reported dipole kick simulations\cite{laxdal1997separated} and is insufficient to cause transmission degradation or aberrations.  
Finally, Figure \ref{fig:ramp} shows projections of the output reference particle $E/A$ for solutions of minimized $\delta P/P$, where $\phi$ has been optimized for each $V$ in the figure. These define the variable $E/A$ states used for operational beam delivery, with an rf limit of $V\sim$6.5\,MV. The inherent advantage of IH structures with an accelerating beta profile design is a better than linear response of the beam energy to the rf amplitude scaling. The output $E/A$ values for the downstream bunching cavity, a triple-gap buncher\cite{Bylinsky:1997fi,Feschenko:2001ed} (Fig.\,\ref{fig:DTL}) are shown in blue, conforming to eq.\,\ref{eq:energyTTF}, showcasing the nonlinear relationship between $E/A$ and $V$ in the IH structure.

\section{Experimental Validation at TRIUMF}

Built during the late 90's as part of the Isotope Separator and Accelerator (ISAC) facility\cite{dilling2014isac}, the post-accelerator at TRIUMF uses an RFQ paired with an 8-cavity 106\,MHz drift tube linac (Fig.\,\ref{fig:DTL}) for beam delivery, driving investigations of nuclear structure and properties. The linac consists of five IH accelerating cavities, where the design $\beta$ changes by more than 10\%, along with three triple-gap split ring resonators\cite{Bylinsky:1997fi}, immediately downstream of Tanks 1-3. A two-gap spiral buncher\cite{mitra199935} upstream of the first IH structure provides longitudinal injection focusing. Transverse focusing is provided by four sets of magnetic quadrupole triplets\cite{laxdal2000first}, interspersed along the lattice. The rf operates\cite{fong2001commissioning} in continuous wave (cw) mode, maximizing time-averaged beam intensity. This {\it separated function}\cite{laxdal1997separated} linac concept gives each of the rf structures their own separate and independently phased amplifier. This design enables fully variable output beam energy per nucleon ($E/A$) in the range\cite{marchetodecelerating,marchetto2008isac} 0.1\,MeV/u\,$\leq$\,$E/A$\,$\leq$\,1.8\,MeV/u, while minimizing emittance growth. Moreover, the mass-to-charge ratio ($A/q$) acceptance of the machine is of 2\,$\leq$\,$A/q$\,$\leq$\,6, enabling charge state selection with a stripping foil, for beam composition purification. Bunches typically exit the RFQ with $\delta P/P$\,$=$\,0.5\% and a half-length of 0.3\,cm and a longitudinal emittance of 15\,$\upmu$m at 153\,keV/u.

The machine has two documented modes of operation, {\it full energy} and {\it variable output energy}. In the former, the linac is configured with each IH structure set to maximum acceleration, while the triple-gap cavities are operated as bunchers, minmizing the longitudinal divergence\cite{Marchetto:2008zz}. In this manner, the output $E/A$ is 1.53\,MeV/u. For variable output energy, the most downstream tank that is powered can vary its contribution to the beam energy from zero to the maximum possible effective voltage that the rf equipment can sustain. If the upstream IH structure's output $\beta(s)$ is lower than the desired final value, the triplet-gap cavity phasing is set to -45$^\circ$, midway between maximum energy gain and fully bunching. The small increase in velocity that is thereby provided reduces the requisite field intensity in the IH structures. In either operational mode, the independence of cavity rf together with the velocity dependent magnetic quadrupole optics causes a large configuration space for machine tuning.


A particularity of the separated function design is the re-configuration of the rf parameters for each resonator necessary to change the output beam energy, when operating in variable output energy mode. In the case of linacs consisting of a series of two-gap resonators such as the TRIUMF SCRF or the linac at FRIB\cite{Ostroumov:2022zrg}, this operation can be predicted by using the transit time method together with a time-of-flight calculation to compute the requisite rf phasing and amplitude. But for the separated function linac this is not possible due to the multiple accelerating solutions of the IH structures in which the synchronous particle velocity changes appreciably. 

{\sc transoptr} permits fast energy change optimizations of an IH cavity linac, by optimizing the rf cavity parameters necessary for acceleration at a user specified output energy. For this, a calibration must first be established between model and machine.


\subsection{Beam-Based Amplitude Calibration\label{sec:calib}}

\begin{figure}[!htpb]
    \centering
    \includegraphics[width=0.48\textwidth]{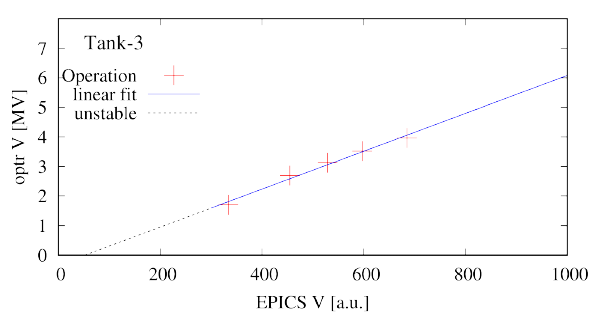}\\
    \caption{TRANSOPTR (optr) model $V$ and {\sc epics} control system rf amplitude, allowing for the computation of a linear calibration between model and machine parameters. DTL Tank-3, the 15-gap cavity from Fig.\,\ref{fig:twofields}, is shown. Computed tune optimizations can thus be loaded directly to the control system.}
    \label{fig:calib}
\end{figure}

Using operational tunes, rf amplitudes and energy readings have been used to calibrate machine and model. Figure \ref{fig:calib} shows the recorded control system\cite{dalesio1991epics} rf amplitudes and corresponding model $V$ necessary to achieve optimum energy gain (minimized $\delta P/P$) for the 15-gap IH structure. This was recorded at a dispersive energy measurement station downstream of the linac, which consists of a 90$^\circ$ horizontal dipole magnet coupled with a vertical parallel wire detector, allowing for a resolution $\Delta E/E$ of 0.1\% and an absolute energy error of 1\%. As the IH cavity phase variation can result in a considerable change in $\beta(s)$, the dipole along with roughly 20 magnetic quadrupoles in both DTL and downstream beamlines must also be tuned in parallel to the cavity, to maintain transmission\cite{marchetto2007radioactive}. 

The calibration should only be regarded as relevant for control system rf amplitude values above roughly 300. In practice, below this value the effects of multipacting can cause instabilities which trip the amplifiers and so operation in this region is avoided. Breakdown at low excitation results in the linear fits not intercepting zero. As a general rule, the cavities are powered-on at approximately $V$\,$=$\,2.0\,${\rm MV}$ to avoid these effects and the associated vacuum response in the room temperature copper resonators. Calibration between model and machine is unique to each rf amplifier. In the event that the equipment is changed, the calibration process must be repeated with beam. Once established, these also allow for the passive monitoring of rf and beam performance; unexpected degradation in agreement between model and machine may be a warning sign of an unexpected condition in either of the beam, cavity of rf equipment.

\subsection{Parameter Optimization}

Individual or groups of parameters can be set for optimization in the envelope code, which will attempt to fit the required values to the constraints using either downhill simplex or simulated annealing\cite{baartman2016transoptr} subroutines. In either case, a vector $\vec{\chi}$ is defined as the sum of the scaled differences between calculated and desired values\cite{TRI-BN-22-08}, and constraints can be applied to either the beam or transfer matrices at any point along the reference trajectory $s$. Using the calibrations from Section \ref{sec:calib}., {\sc transoptr} is used to compute the necessary ($\phi,V$) values for the IH cavities in the ISAC-DTL (Figure \ref{fig:DTL}). In this case, by restricting the optimizer to a single rf cavity and supplying a desired output $E/A$, only two parameters need to be optimized at a time. The on-axis energy gain of the reference particle is found using the computed time coordinate $t_0(s)$, solved at each numerical integration step along the reference trajectory. Additionally, exploiting the condition that IH cavities, for any attainable output energy, are operated at minimized longitudinal momentum spread, suggests the constraint:\begin{equation}
    \sigma_{66} = \langle \delta P^2\rangle \longrightarrow 0.
    \label{eq:sigma66zero}
\end{equation} 
However, together with the output $E/A$, this produces two constraints for two fit parameters, insufficient to find the optimum ($\phi,V$), given the complex behavior of multigap cavities, such as Tank-3 (Fig.\,\ref{fig:topo}, bottom). Instead, the condition (\ref{eq:sigma66zero}) can be augmented by including constraints upon the transverse optics inherent to the rf accelerating field, by requiring the transfer matrix through the IH cavity conform to:
\begin{equation}
    {\mathbf M_{21}} + {\mathbf M_{43}} + {\mathbf M_{65}} = 2{\mathbf M_{21}} + {\mathbf M_{65}} \longrightarrow 0,
    \label{eq:constraint}
\end{equation}
which minimizes the accumulated transverse rf focal effects through the field\cite{shelbaya2021autofocus} while providing a small amount of bunching, preventing excessive longitudinal growth while also keeping $\delta P$ minimized. This enables the imposition of 4 constraints, three on the transfer matrix and one on the reference particle $\beta(s)$, for the obtention of two rf parameters, improving the performance for the optimization. 

\begin{figure}[!htpb]
    \centering
    \includegraphics[width=0.48\textwidth]
    {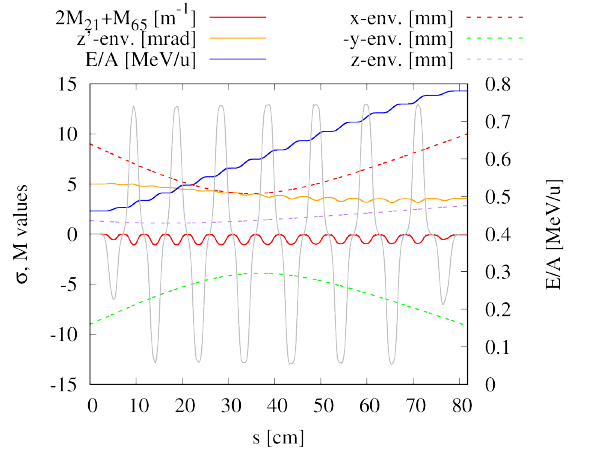}\\
    \caption{\label{fig:optimizeTank3}{\sc transoptr} computed beam transfer matrix and 2rms containment envelopes along with the reference particle $E/A$ through the 15-gap IH Tank-3 of the ISAC-DTL, for an $^{23}$Na$^{6+}$ beam. The arbitrarily normalized intensity of the on-axis electric field $\mathcal{E}(s)$ for the cavity is shown in grey for reference.}
\end{figure}

\begin{figure}[!t]
    \centering
    \includegraphics[width=0.46\textwidth]{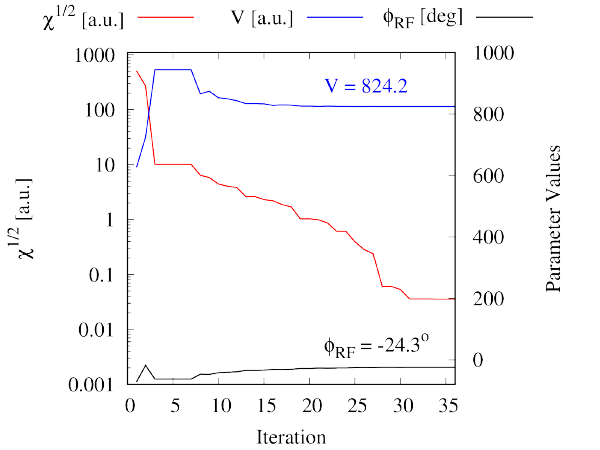}
    \caption{\label{fig:minimization}{\sc transoptr} optimization of rf parameters ($\phi,V$) for the 15-gap accelerating cavity, using simulated annealing, constrained for a reference particle $E/A$\,$=$\,0.781\,MeV/u and the condition of eq.\,\ref{eq:constraint}. The obtained solution corresponds to what is shown in Figure \ref{fig:optimizeTank3}.}
\end{figure}

This has been applied to the 15-gap field (Fig.\,\ref{fig:twofields}, middle), where Figure \ref{fig:optimizeTank3} shows the beam envelopes and reference particle $E/A$ during acceleration. At this optimized energy gain, the transverse rf focusing/defocusing is minimized, and the IH structure's transverse optics approach that of a drift in free space. Incidentally, this allows for the computation of the transverse tune through optimally configured IH structures using a discrete step in the energy at the mid-point, surrounded by drifts through free space. {\sc transoptr}'s convergence toward the final rf parameters, using simulated annealing, is shown in Figure \ref{fig:minimization}. The optimum solution is found after 36 iterations, with the optimization taking 0.8\,sec on a conventional pc. This allows for the fast evaluation of the full 6-D envelopes through such structures. The method is general, provided knowledge of the on-axis electric field amplitude $\mathcal{E}(s)$, obtained either by rf measurements or using simulation software such as {\sc opera2D}\cite{fields1999opera} or {\sc cst-mws}. This has been developed into an energy change optimization algorithm, used to find the optimum rf parameters for each IH structure, for a user defined output energy.
\subsection{Model-Coupled Energy Changes}
For a variety of reasons, at TRIUMF-ISAC the cavity rf phases can vary from time to time\cite{TRI-BN-17-24,TRI-BN-20-19}, rendering phase calibrations impractical. Instead, by reading-in the real-time rf scaling $V$ and knowing that the IH structures are always configured per the condition in eq.\,\ref{eq:constraint}, parameter optimizations for IH cavity energy changes can be performed, without requiring a phase calibration. 

Model optimizations are now being used for on-line linac tuning, allowing for the computation of cavity ($\phi,V$) needed to achieve any desired beam velocity, with {\sc transoptr}. Web-based communication with the accelerator control system variables for the DTL rf amplitudes is made possible using an in-house server based remote reading service, which broadcasts the real-time values to a standardized http address. A database stores the real-time rf parameters. A script monitors the control system values, triggering an update if any changes are detected, ensuring the model always uses the most up to date values. Software has been written to automatically generate\cite{TRI-BN-22-06} and execute the {\sc transoptr} simulation files for DTL IH cavity optimization. For any DTL tank, the user must only specify the input and desired output $E/A$, along with beam parameters such as mass and charge. 
\begin{figure}[!b]
    \centering
      \includegraphics[width=0.45\textwidth]{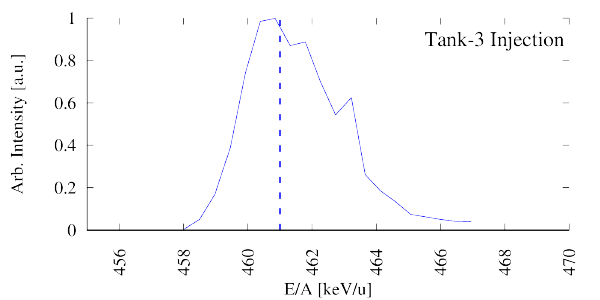}\\
      \includegraphics[width=0.45\textwidth]{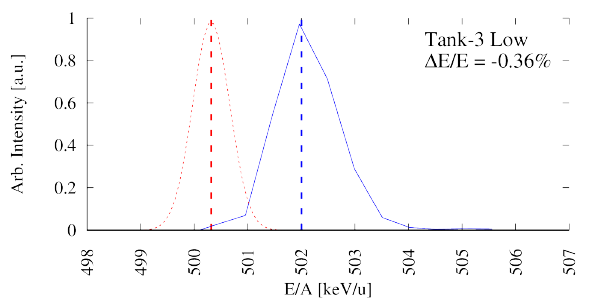}\\
      \includegraphics[width=0.45\textwidth]{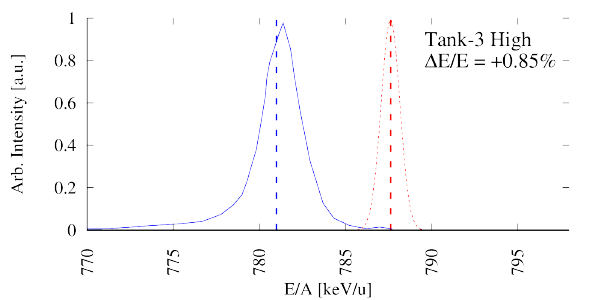}
    \caption{\label{fig:DTLDE}{\sc TRANSOPTR} energy prediction for a $^{23}$Na$^{6+}$ beam, compared to on-line measurements. The dotted red lines show the model predicted energies for the bunch center to first order, based on the amplitude calibration of Fig.\,\ref{fig:calib}. The blue distributions show dispersive energy measurements performed at the ISAC energy diagnostic station. The model prediction for the energy spectra are displayed in red using Gaussians, based on the longitudinal tune in\cite{shelbaya2021autofocus} for comparison.}
\end{figure}In turn, the cavities are set to the model computed optima, allowing the operator to directly set the IH cavity, instead of ramping the tank while attempting to maintain high transmission. This constraint, together with the rf amplitude calibration is sufficient to find the ($\phi,V$) pair for any $\beta(s)$ within an IH cavity's operating range. The caveat remains that the first order envelope code assumes the bunch center and reference particles are one and the same. 

Operationally, this is sufficient to predict the IH structure acceleration energies to about 1\% accuracy. This is because the machine is not configured following a pure KONUS tune for beam delivery, but rather for longitudinal energy spread minimization. Knowing the RF amplitudes alone, the model can be used to perform a phase optimization, returning a predicted output energy for each tank. Figure \ref{fig:DTLDE} shows the algorithm's prediction of the on-line beam energy for the 15-gap IH Tank-3, compared to beam-based measurements for $^{23}$Na$^{6+}$ on the ISAC energy diagnostic.

\section{Summary}

Transit time factor treatment of beam energy gain is colloquially referred to in accelerator physics as $V\cos\phi$, suitable for the computation of the output reference particle energy in a small energy change, where the synchronous velocity remains nearly constant across the structure. For circular machines, the trajectory enables many successive rf gap transits, allowing for the ramping of the beam energy to considerable values. However, for most linear accelerators only a single pass is possible. As the gap count grows and the structure's spatial period varies in length, the TTF approach becomes more involved. This allows large (desirable) changes to the reference particle velocity, though the price to pay is the inapplicability of a first order energy change approximation.

A Hamiltonian approach, in which the centroids and second moments of the distribution are tracked through the time-varying fields to first order, enables a computationally lightweight means for the analysis of multigap accelerated beam dynamics. {\it Caveat emptor:} the linear optics requires an understanding of the region of applicability and validity of the first order treatment. This avoids the intricacies of multigap TTF approaches. 

In a first instance, accelerated beam properties through a bunching rf cavity were presented. A brief exploration of small energy change dynamics supported by {\sc transoptr} parameter scans produced insight into the nature of variable output energy operation for two-gap rf resonators. This highlighted the ability to perform single-parameter, continuous energy changes, by way of varying the rf phase and suitably adjusting the rf amplitude per eq.~(\ref{eq:debunch2gap}), to minimize the longitudinal beam divergence at the cavity exit. In other words, the rf parameters for variable output energy operation of small energy change cavities can be analytically predicted.

Next, {\sc transoptr} was used to showcase the more complex beam dynamics of a 15-gap IH cavity which is quasi-periodic; the cell length of the structure is elongated from one gap to the next, enabling the considerable variation of the reference particle velocity. Here, the first order energy change approximation no longer applies and the requisite rf cavity phase and voltage amplitude necessary for variable output energy acceleration must be solved for numerically. Phase slippage in the rf cavity causes the relationship between the optimum phase and rf voltage to become much more complex than for the small energy change case of eq.\,(\ref{eq:debunch2gap}). The computationally lightweight first order envelope method discussed herein produces fast computation times, on the order of one second for a full optimization on a conventional pc.

Finally, using the {\sc transoptr} model of the rf cavity, which is entirely represented by its on-axis longitudinal electric field, a calibration has been established with beam, between model and machine. {\sc transoptr}'s numerical optimization capability is used to solve for the optimum energy gain through IH cavities, minimizing certain transfer matrix elements through the field. The predictions are found to be accurate to roughly 1\% or less when compared to on-line readings at the variable energy TRIUMF ISAC-DTL. Reducing the uncertainty on the energy analyzing magnet's field could improve the accuracy of the method. The presented methodology is being used to develop accelerator tuning software at TRIUMF which aims to use {\sc transoptr} as part of a real-time digital twin of the machine, enabling operators to perform a variety of model-coupled interventions on the linac. This also enables machine learning investigations, presently being pursued\cite{wangaccelerator}, now capable of including and optimizing multigap cavities.

\section{Data Availability}

The data that support the findings of this study are available from the corresponding author upon reasonable request. 

\section{Acknowledgements}

Thanks to R.\,Laxdal and M.\,Marchetto for useful discussions. J.\,Aoki and the Operations group are thanked for their assistance. S.\,Kiy is thanked for helpful discussions. Gratitude to R.\,Leewe and the RF group members for fruitful discussions. TRIUMF is funded under a contribution agreement with NRC (National Research Council Canada.) The TRIUMF campus is located on the traditional, ancestral, and unceded territory of the Musqueam people.\vfill


\begin{thebibliography}{10}

\bibitem{kester2003accelerated}
O.~Kester, T.~Sieber, S.~Emhofer, F.~Ames, K.~Reisinger, P.~Reiter, P.~G. Thirolf, R.~Lutter, D.~Habs, B.~H Wolf, et~al.,
\newblock {Accelerated radioactive beams from REX-ISOLDE},
\newblock Nuclear Instruments and Methods in Physics Research Section B: Beam Interactions with Materials and Atoms {\bf 204}, 20 (2003).

\bibitem{Barth:2000bf}
W.~Barth, P.~Forck, J.~Glatz, W.~Gutowski, G.~Hutter, J.~Klabunde, R.~Schwedhelm, P.~Strehl, W.~Vinzenz, D.~Wilms, et~al.,
\newblock {Commissioning of IH-RFQ and IH-DTL for the GSI High-Current LINAC},
\newblock in {\em Proc. LINAC'00}, number~20 in Linear Accelerator Conference, pages 229--231, JACoW Publishing, Geneva, Switzerland (2000).

\bibitem{ball2020triumf}
G.~Ball, I.~Dillmann, A.~Garnsworthy, G.~Gwinner, R.~Kanungo, G.~Morris, Gerald, C.~Ruiz,
\newblock {The TRIUMF-ISAC Facility: Recent Highlights in RIB Science and Future Prospects with ARIEL},
\newblock {\em Nuclear Physics News} {\bf 30}, pages 27--32 (2020).

\bibitem{blewett1956linear}
J.~Blewett,
\newblock {Linear Accelerator Injectors for Proton Synchrotrons},
\newblock in {\em CERN Symposium on High-Energy Accelerators and Pion Physics}, pages 159--166 (1956).

\bibitem{Gerigk:2011ph}
F.~Gerigk,
\newblock {Cavity types},
\newblock in {\em CERN Accelerator School: RF for Accelerators} (2011).

\bibitem{laxdal1997separated}
R.~Laxdal, P.~Bricault, T.~Reis, and D.~Gorelov,
\newblock A Separated Function Drift-Tube Linac for the ISAC Project at TRIUMF,
\newblock in {\em Proc. PAC'97},
  IEEE, volume~1, pages 1194--1196, Vancouver, BC, Canada (1997).

\bibitem{ratzinger2019combined}
U.~Ratzinger, H.~H{\"a}hnel, R.~Tiede, J.~Kaiser, and A.~Almomani,
\newblock {Combined Zero Degree Structure Beam Dynamics and Applications},
\newblock {\em Physical Review Accelerators and Beams} {\bf 22}, 114801 (2019).

\bibitem{Xiao:2019xlq}
C.~Xiao, X.~Du, and L.~Groening,
\newblock {Investigations on KONUS Beam Dynamics Using the Pre-Stripper Drift Tube LINAC at GSI},
\newblock in {\em {14th International Conference on Heavy Ion Accelerator Technology}}, page WEOAA02 (2019).

\bibitem{panofsky1951linear}
W.~K.~H. Panofsky,
\newblock {Linear Accelerator Beam Dynamics}, {\em Technical Report URCL-1216,}
\newblock {University of California Radiation Laboratory} (1951).

\bibitem{Shelbaya:2021sth}
O.~Shelbaya, R.~A. Baartman, and O.~K. Kester,
\newblock {End-to-End RMS Envelope Model of the ISAC-I Linac},
\newblock in {\em Proc. IPAC'21}, pages 4183--4186, JACoW Publishing, Geneva, Switzerland (2021).

\bibitem{shelbaya2019fast}
O.~Shelbaya, R.~Baartman, and O.~Kester,
\newblock {Fast radio frequency quadrupole envelope computation for model based beam tuning,}
\newblock {\em Physical Review Accelerators and Beams} {\bf 22}, 114602 (2019).

\bibitem{Dutto:2001wy}
G.~Dutto, K.~Fong, R.~Laxdal, G.~Mackenzie, M.~Pasini, R.~Poirier and R.~Ruegg,
\newblock {Beam Commissioning and First Operation of the ISAC DTL at TRIUMF},
\newblock in {\em Proc. PAC'01}, pages 3942--3944, JACoW Publishing, Geneva, Switzerland (2001).

\bibitem{Laxdal:2000hx}
R.~E.~Laxdal, P.~Bricault, G.~Dutto, K.~Fong, G.~H.~Mackenzie, R.~Poirier, W.~Rawnsley, R.~Ruegg and G.~Stinson,
\newblock {First Beam Test with the ISAC Separated Function DTL},
\newblock in {\em Proc. LINAC'00}, pages 244--246, JACoW Publishing, Geneva, Switzerland (2000).

\bibitem{Wangler:2008zz}
T.~P. Wangler,
\newblock {\em {RF linear accelerators}},
\newblock John Wiley \& Sons (2008).

\bibitem{Lallement:2017uxp}
J.-B. Lallement,
\newblock {Experience with the Construction and Commissioning of Linac4},
\newblock in {\em {28th International Linear Accelerator Conference}}, page TU1A03 (2017).

\bibitem{shelbaya2021autofocus}
O.~Shelbaya, T.~Angus, R.~Baartman, P.~M.~Jung, O.~Kester, S.~Kiy, T.~Planche, S.D.~R\"adel,
\newblock Autofocusing Drift Tube Linac Envelopes,
\newblock {\em Physical Review Accelerators and Beams} {\bf 24}, 124602 (2021).

\bibitem{Marchetto:2008zz}
M.~Marchetto, J.~Berring, and R.~E. Laxdal,
\newblock {Upgrade of the ISAC DTL Tuning Procedure at TRIUMF},
\newblock in {\em Proc. EPAC'08}, number~11 in European Particle Accelerator Conference, pages 3440--3442, JACoW Publishing, Geneva, Switzerland (2008).

\bibitem{Baartman:2017faa}
R.~Baartman,
\newblock {Fast Envelope Tracking for Space Charge Dominated Injectors},
\newblock in {\em {28th International Linear Accelerator Conference}}, pages 1017-1021, JACoW Publishing, Geneva, Switzerland (2017).

\bibitem{Courant:1958wbj}
E.~D. Courant and H.~S. Snyder,
\newblock {Theory of the Alternating Gradient Synchrotron},
\newblock {\em Annals of Physics} {\bf 3}, 1 (1958).

\bibitem{brown1967slac}
K.~L. Brown,
\newblock {{A First and Second Order Matrix Theory for the Design of Beam Transport Systems and Charged Particle Spectrometers}},
\newblock {\em Adv. Part. Phys.}, pages 71-134, also {\em internal report SLAC-75} {\bf 1}, 71 (1968).

\bibitem{sacherer1971rms}
F.~Sacherer,
\newblock {{RMS} Envelope Equations with Space Charge},
\newblock {\em {IEEE} Transactions on Nuclear Science} {\bf 18}, 1105 (1971).

\bibitem{heighway1981transoptr}
E.~Heighway and R.~Hutcheon,
\newblock {{{TRANSOPTR}—A Second Order Beam Transport Design Code with Optimization and Constraints}},
\newblock {\em Nuclear Instruments and Methods in Physics Research} {\bf 187}, 89 (1981).

\bibitem{baartman2016transoptr}
R.~Baartman,
\newblock TRANSOPTR: Changes Since 1984, 
\newblock {\em Technical Report TRI-BN-16-06}, TRIUMF (2016).

\bibitem{Shelbaya:2022eyc}
O.~Shelbaya, R.~Baartman, O.~Kester, S.~Kiy, and S.~R\"adel,
\newblock Model Coupled Accelerator Tuning With an Envelope Code, {\em Proc. LINAC'22}, pages 549-551, JACoW Publishing, Geneva, Switzerland (2022).


\bibitem{mitra199935}
A.~Mitra and R.~Poirier,
\newblock {A 35 MHz Spiral Re-Buncher Cavity for the TRIUMF ISAC Facility},
\newblock in {\em Proc. PAC'99},
  volume~2, pages 839--841, IEEE (1999).

\bibitem{shelbaya2020transoptr}
O.~Shelbaya,
\newblock {The TRANSOPTR Model of the ISAC Drift Tube Linear Accelerator - Part I: Longitudinal Verification},
\newblock {\em Technical Report TRI-BN-20-08}, TRIUMF (2020).

\bibitem{TRI-BN-19-02}
O.~Shelbaya,
\newblock {TRANSOPTR Implementation of the MEBT Beamline},
\newblock {\em Technical Report TRI-BN-19-02}, TRIUMF (2019).

\bibitem{fields1999opera}
V.~Fields,
\newblock {Opera-2d user guide},
\newblock Vector Fields Limited, England (1999).

\bibitem{studio2008cst}
{CST-MWS},
\newblock CST Studio Suite (CP Studio, Darmstadt, Germany) (2008).

\bibitem{Bylinsky:1997fi}
Y.~Bylinsky, V.~Kukhtiev, P.~Ostroumov, V.~Paramonov, and R.~E. Laxdal,
\newblock {A Triple Gap Resonator Design for the Separated Function DTL at TRIUMF},
\newblock in {\em Proc. PAC'97}, pages 1135--1137, JACoW Publishing, Geneva, Switzerland (1997).

\bibitem{Feschenko:2001ed}
A.~Vasyuchenko, A.~Feschenko, A.~Kvasha, A.~Menshov, V.~Paramonov, Y.~Bylinsky, G.~Dutto, R.~E. Laxdal, A.~K. Mitra, R.~Poirier,
\newblock {Development, Fabrication and Test of Triple Gap Split-Ring Bunchers for the TRIUMF ISAC Facility},
\newblock in {\em Proc. PAC'01}, pages 978--980, JACoW Publishing, Geneva, Switzerland (2001).

\bibitem{dilling2014isac}
J.~Dilling, R.~Kr{\"u}cken, and G.~Ball,
\newblock {\em Hyperfine Interactions} {\bf 225}, 1 (2014).

\bibitem{laxdal2000first}
R.~E. Laxdal, P.~Bricault, G.~Dutto, K.~Fong,G.~H. Mackenzie, R.~Poirier, W.~Rawnsley,
\newblock First beam test with the ISAC Separated Function DTL,
\newblock in {\em Proc. 2000 Linac Conference} (2000).

\bibitem{fong2001commissioning}
K.~Fong, S.~Fang, M.~Laverty, J.~Lu, and L.~Poirier,
\newblock {Commissioning of the TRIUMF ISAC RF System},
\newblock in {\em Proc. PAC'01}, volume~2, pages 945--947, IEEE (2001).

\bibitem{marchetodecelerating}
M.~Marchetto, R.~E. Laxdal, and F.~Yan,
\newblock {Decelerating Heavy Ion Beams Using the ISAC DTL},
\newblock in {\em Proc. 11th Int. Conf. on Heavy Ion Accelerator Technology
  (HIAT'09)}, pages 261--265, JACoW Publishing (2009).

\bibitem{marchetto2008isac}
M.~Marchetto,
\newblock {ISAC-II Operation and Future Plans},
\newblock in {\em Proc. LINAC'08}, number~24 in Linear Accelerator Conference, pages 1--5, JACoW Publishing, Geneva, Switzerland (2008).

\bibitem{Ostroumov:2022zrg}
P.~N. Ostroumov, F.~Casagrande, K.~Fukushima, M.~Ikegami, T.~Kanemura, S.~Kim, S.~Lidia, G.~Machicoane, T.~Maruta, D.~Morris, et~al.,
\newblock {Status of FRIB commissioning},
\newblock in {\em Proc. 64th ICFA Advanced Beam Dynamics Workshop on High-Intensity and High-Brightness Hadron Beams (HB 2021)}, pages 203--207 (2021).

\bibitem{dalesio1991epics}
L.~R. Dalesio, A.~Kozubal, and M.~Kraimer,
\newblock EPICS architecture,
\newblock {\em Technical report LA-UR-91-3543}, Los Alamos National Lab., NM (United States) (1991).

\bibitem{marchetto2007radioactive}
M.~Marchetto, Z.T.~Ang, K.~Jayamanna, R.E.~Laxdal, A.~Mitra, V.~Zvyagintsev,
\newblock {\em The European Physical Journal Special Topics} {\bf 150}, 241 (2007).

\bibitem{TRI-BN-22-08}
R.~Baartman,
\newblock {TRANSOPTR Reference Manual},
\newblock {\em Technical Report TRI-BN-22-08}, TRIUMF (2022).

\bibitem{TRI-BN-17-24}
O.~Shelbaya,
\newblock {Preliminary Investigation of ISAC-I RF/Temperature Correlations and General Performance},
\newblock {\em Technical Report TRI-BN-17-24}, TRIUMF (2017).

\bibitem{TRI-BN-20-19}
S.~Kiy,
\newblock {LLRF Phase Shifter Calibrations},
\newblock {\em Technical Report TRI-BN-20-19}, TRIUMF (2020).

\bibitem{TRI-BN-22-06}
O.~Shelbaya and P.~M.~Jung,
\newblock {Generation of TRANSOPTR files with xml2optr},
\newblock {\em Technical Report TRI-BN-22-06}, TRIUMF (2022).

\bibitem{wangaccelerator}
D.~Y. Wang, H.~Bagri, C.~Macdonald, S.~Kiy, P.~M.~Jung,  O.~Shelbaya, T.~Planche, W.~Fedorko, R.~Baartman, O.~Kester,\newblock {Accelerator Tuning with Deep Reinforcement Learning},
\newblock in {\em Workshop at the 35th Conference on Neural Information Processing Systems}, Vancouver, BC, Canada (2021).

\end{thebibliography}
\providecommand{\noopsort}[1]{}\providecommand{\singleletter}[1]{#1}%

\end{document}